\documentclass[12pt]{article}
\def\be{\begin{equation}}
\def\ee{\end{equation}}
\def\bea{\begin{eqnarray}}
\def\eea{\end{eqnarray}}
\def\ba{\begin{array}}
\def\ea{\end{array}}

\begin{document}

\title{Chiral quark model ($\chi$QM) and the nucleon spin}
\author{Harleen Dahiya and Manmohan Gupta    \\
{\it Department of Physics,} \\
{\it Centre of Advanced Study in Physics,} \\
{\it Panjab University, Chandigarh-160 014, India.}}
\maketitle
 
\begin{abstract}
Using $\chi$QM with configuration mixing, the contribution of the 
gluon polarization to the flavor singlet component of the total spin 
has been calculated phenomenologically 
through the relation $\Delta\Sigma(Q^2)=\Delta\Sigma-
\frac{3\alpha_s(Q^2)}{2\pi}\Delta g(Q^2)$ as defined in the Adler-Bardeen 
scheme, where $\Delta\Sigma$ on the right hand side is $Q^2$ independent. 
For evaluation the contribution of gluon polarization 
$\Delta g'(=\frac{3\alpha_s(Q^2)}{2\pi}\Delta g(Q^2)$), $\Delta\Sigma$ 
is found in the $\chi$QM by fixing the latest E866 data pertaining
to $\bar u-\bar d$ asymmetry and the spin polarization functions
whereas  $\Delta \Sigma(Q^2)$ is taken to be $0.30\pm0.06$ and
$\alpha_s= 0.287\pm0.020$, both at $Q^2=5{\rm GeV}^2$.
The contribution of gluon polarization $\Delta g'$
comes out to be $0.33$ which 
leads to an almost perfect fit for spin distribution functions in the 
$\chi$QM. When its implications for magnetic moments are investigated, 
we find perfect fit for many of the magnetic moments. 
If an attempt is made to explain the angular
momentum sum rule for proton by using the above value of $\Delta g'$, 
one finds the contribution of gluon angular momentum to be as important 
as that of the $q \bar q$ pairs.
\end{abstract}

Ever since the measurements of
polarized structure functions of proton in the deep inelastic scattering 
(DIS) experiments \cite{EMC,adams,abe}, showing that the valence 
quarks of the proton carry only about 30\% of its spin, several 
interesting facts have been revealed regarding the polarized structure 
functions of the nucleon as well as the quark distribution functions in 
these experiments. The present experimental situation 
\cite{{adams},{abe},{PDG}} in terms
of the polarized structure functions, $\Delta u$, $\Delta d$ and $\Delta s$, 
measuring the spin polarizations is summarized as follows:
\bea
\Delta u& =& 0.85\pm 0.05~ \cite{adams}, ~~~ 
\Delta d = -0.41 \pm .05~  \cite{adams}, ~~~ 
\Delta s = -0.07 \pm 0.05~ \cite{adams}\,, \label{ds} \\
\Delta \Sigma &=&0.30\pm 0.06~ \cite{abe},~~~
\Delta_3= 1.267 \pm .0025~ \cite{PDG}, ~~~
\Delta_8= 0.58 \pm 0.025~ \cite {PDG} \label{d8}\,,
\eea
where $\Delta \Sigma$ corresponds to the flavor 
singlet component and $\Delta_3$,  $\Delta_8$ correspond to the flavor 
non-singlet components of the total spin.
Further, the experimental measurements suggest absence of the 
polarizations of the antiquarks \cite{adams}.

Similarly, DIS experiments have
given fairly good deal of information about the quark distribution
functions, for example, the  $\bar u-\bar d$  asymmetry or $\bar u/\bar d$ 
ratio is given as follows
\bea 
{\bar u-\bar d}&=& -0.147 \pm 0.024~ \cite{NMC}, ~~~
{\bar u/\bar d}=0.51 \pm 0.09~ \cite{baldit}\,, \\
{\bar u-\bar d}&=& -0.118 \pm 0.018~ \cite{E866}, ~~~
{\bar u/\bar d}= 0.67 \pm 0.06~ \cite{E866}\,. \label{antiquark}
\eea
The E866 experiment \cite{E866} provides by far the best measurement 
indicating that the nucleon sea contains more
 number of $\bar d$ quarks than the $\bar u$ quarks.  

The measured spin polarizations as well as the quark distribution
functions can be related to certain well known sum rules such as
Bjorken sum rule (BSR) \cite{bjorken}, 
Ellis-Jaffe sum rule (EJSR) \cite{ellis} and the
Gottfried sum rule (GSR) \cite{GSR}. These sum rules can be derived within QCD
using operator product expansion, renormalization group invariance and 
isospin conservation in the DIS. Further, these sum rules, having weak
$Q^2$ dependence \cite{ellisq2}, can be related to certain low energy 
parameters hence providing vital clues 
to the dynamics of the low energy regime or nonperturbative regime of QCD.

The spin distribution functions $\Delta_3$
and $\Delta_8$, expressed through Eq. (\ref{d8}),
can be related to BSR \cite{bjorken} and the EJSR \cite{ellis} as
\bea
BSR: ~~~~~~\Delta_3&=& \Delta u-\Delta d=F+D\,, \label{BSR}\\
EJSR:~~~~~~\Delta_8&=& \Delta u+\Delta d-2 \Delta s=3F-D\,, \label{EJSR}
\eea
where $F$ and $D$ are the well known parameters measured in the 
neutron $\beta-$decay and the weak decays of hyperons.
Similarly, the $\bar u-\bar d$ asymmetry is related to Gottfried sum rule 
violation \cite{GSR}, for example,
\be   
\left[I_G-\frac{1}{3}\right ] =
\frac{2}{3} (\bar u-\bar d)\,,  \label{gsr}
\ee 
where $I_G=\int_{0}^{1}dx\frac{F_2^p(x)-F_2^n(x)}{x}$ is the Gottfried 
integral.

On the other hand, non relativistic quark model (NRQM),
quite successful in explaining a good deal of low energy data 
\cite{DGG,Isgur,yaouanc,photo}, 
has the following predictions for the above mentioned quantities
\bea
\Delta u &=& 1.33, ~~~\Delta d = -0.33, ~~~ \Delta s = 0\,,  \\
\bar u-\bar d&=& 0,~~~
\Delta \Sigma= 1, ~~~\Delta_3= 1.66,  ~~~\Delta_8= 1\,.
\eea
One immediately finds that the NRQM predictions are in considerable
disagreement with the above mentioned DIS measurements. 
The disagreement between the NRQM spin
polarization predictions and the DIS measurements can broadly be
characterized as ``proton spin crisis''.
Apart from the above mentioned difficulties which one faces in explaining
the DIS data, to have a deeper understanding of 
the deep inelastic results as well as the dynamics of the constituents 
of the nucleon, one also has to explain the ``angular momentum sum rule'' 
\cite{jaffe} which is expressed as
\be
\frac {1}{2} =\frac{1}{2}\Delta \Sigma+\Delta L_{q}+\Delta G+
\Delta L_g\,, \label{total spin}
\ee
where $\Delta \Sigma$ is the spin polarization contribution of
the quarks, $\Delta L_{q}$ is the orbital 
angular momentum of the quarks, $\Delta G$ is the 
gluon polarization and $\Delta L_g$ is the orbital angular 
momentum of the gluons. Recently, some efforts have been made 
in the chiral quark model 
\cite{tommy} and chiral quark soliton model \cite{cqsm} to 
understand the above sum rule.
However, a detailed understanding of the partitioning of nucleon spin
expressed through the sum rule does not seem to have been achieved, 
in particular, at present we do not have any clue about the likely
magnitude of $\Delta L_g$ either from the experiments or theoretical models.
An understanding of the experimental information, expressed through the
Eqs. (\ref{ds})-(\ref{antiquark}) as well as the spin partition of the
 nucleon expressed 
through Eq. (\ref{total spin}), constitute a major challenge for any model 
trying to explain the nonperturbative regime of QCD. 

In this context, the $\chi$QM as formulated by Manohar and Georgi 
\cite{{manohar},{wein}}, has recently got good deal of
attention \cite{{eichten},{cheng},{song},{johan}}
as it not only provides a viable description of depolarization
of valence quarks through the emission of a Goldstone boson (GB)
which causes a modification of the flavor content but is also able to
account for the $\bar u-\bar d$ asymmetry
{\cite{{NMC},{E866},{GSR}}}, existence of significant strange 
quark content $\bar s$ in the nucleon,  various
quark flavor contributions to the proton spin {\cite{eichten}}, baryon
magnetic moments {\cite{{eichten},{cheng}}}, absence of polarizations of 
the antiquark sea  in the nucleon \cite{cheng,song,antiquark} and
hyperon $\beta-$decay parameters etc..

Recently, it has been shown that invoking  configuration mixing,
having its origin in spin-spin forces, in the $\chi$QM (referred to 
as $\chi$QM$_{gcm}$) with SU(3) and
axial U(1) symmetry breakings improves the predictions of $\chi$QM
regarding the spin and quark distribution functions
\cite{hd,hdorbit,su4}. It has also been shown in a very recent
communication \cite{hdorbit} that when orbital angular momentum of the
sea quarks  is taken into
account through the Cheng-Li mechanism \cite{cheng1}, $\chi$QM$_{gcm}$
is able to give an excellent fit to magnetic moments with an almost
perfect fit for the violation of Coleman-Glashow sum rule.
In this context, it may be desirable to mention that chiral quark 
soliton model \cite{cqsm}, with quarks and $q \bar q$ pairs as effective
degrees of freedom, when endowed with angular momentum
of the $q \bar q$ pairs is also able to give a fairly good
description of the spin distribution functions of the nucleon.
The success of $\chi$QM in resolving the ``proton spin crisis''
alongwith the above mentioned successes in explaining large amount of
experimental data strongly indicates that constituent quarks,  
weakly interacting GBs and $q \bar q$ pairs alongwith the weakly 
interacting gluons ({\it a la} Manohar and Georgi)
provide the appropriate degrees of freedom at the leading order
in the nonperturbative regime of QCD. 
This then raises the question to what extent
one can understand the partitioning of the nucleon spin within the
basic premises of  $\chi$QM by including the contributions from gluon
polarization.

Further, Kabir and Song in a recent interesting work \cite{kabir}
have found in
the case of $\beta-$decay transitions of the hyperons that the
fit  obtained in terms of the coupling parameters $F$ and $D$ remains
satisfactory even in the presence of SU(3) breaking. This then allows
the fine tuning of the $\chi$QM 
parameters by fitting to $\Delta_3=F+D$ and $\Delta_8=3F-D$ which 
are rather well known in terms of quark spin polarization functions.

In the light of the above developments, on the one hand, it seems
desirable to carry out a fine grained analysis  of $\chi$QM with
configuration mixing by fitting the latest E866 data with the purpose
of fine tuning the $\chi$QM parameters. On the other hand, having
fixed the $\chi$QM parameters, one can then examine closely its
implications for the partitioning of the nucleon spin, in particular,
one would like to phenomenologically estimate the contributions of
gluon polarization at the non-leading order to the flavor singlet
component as well as the gluon angular momentum.

The details of  $\chi$QM$_{gcm}$ have already been discussed in
Ref. \cite{{hd},{hdorbit},su4}, however to facilitate the discussion as
well as readability of the manuscript, some essential details of
$\chi$QM with configuration mixing have been presented in the
sequel. Before proceeding further it needs to be mentioned that we have
adopted the Manohar-Georgi formulation of the $\chi$QM
as extended by Cheng and Li  \cite{cheng1} to ``proton spin crisis'', 
wherein the effective degrees of freedom are constituent quarks, 
weakly interacting
GBs as well as gluons interacting with a ``much smaller'' coupling
constant. 

The basic process in the $\chi$QM is the
emission of a GB by a constituent quark which further splits into a $q
\bar q$ pair, for example,  
\be
  q_{\pm} \rightarrow {\rm GB}^{0}
  + q^{'}_{\mp} \rightarrow  (q \bar q^{'})
  +q_{\mp}^{'}\,, \label{basic}
\ee
where $q \bar q^{'}  +q^{'}$ constitute the ``quark sea''
  \cite{{cheng},{song},{johan},{cheng1}}. 
The effective Lagrangian describing interaction between quarks and a nonet of
GBs, consisting of octet and a singlet, can be expressed as
\be
{\cal L} = g_8 \bar q \Phi q\,,
\ee
\bea 
q =\left( \ba{c} u \\ d \\ s \ea \right),&
~~~~~& \Phi = \left( \ba{ccc} \frac{\pi^0}{\sqrt 2}
+\beta\frac{\eta}{\sqrt 6}+\zeta\frac{\eta^{'}}{\sqrt 3} & \pi^+
  & \alpha K^+   \\
\pi^- & -\frac{\pi^0}{\sqrt 2} +\beta \frac{\eta}{\sqrt 6}
+\zeta\frac{\eta^{'}}{\sqrt 3}  &  \alpha K^0  \\
 \alpha K^-  &  \alpha \bar{K}^0  &  -\beta \frac{2\eta}{\sqrt 6}
 +\zeta\frac{\eta^{'}}{\sqrt 3} \ea \right), \eea
where $\zeta=g_1/g_8$, $g_1$ and $g_8$ are the coupling constants for the 
singlet and octet GBs, respectively.

SU(3) symmetry breaking is introduced by considering
$M_s > M_{u,d}$ as well as by considering
the masses of GBs to be nondegenerate
 $(M_{K,\eta} > M_{\pi})$ {\cite{{song},{johan},{cheng1}}}, whereas 
  the axial U(1) breaking is introduced by $M_{\eta^{'}} > M_{K,\eta}$
{\cite{{cheng},{song},{johan},{cheng1}}}.
The parameter $a(=|g_8|^2$) denotes the transition probability
of chiral fluctuation
of the splittings  $u(d) \rightarrow d(u) + \pi^{+(-)}$, whereas 
$\alpha^2 a$, $\beta^2 a$ and $\zeta^2 a$ respectively 
denote the probabilities of transitions of
$u(d) \rightarrow s  + K^{-(0)}$, $u(d,s) \rightarrow u(d,s) + \eta$,
 and $u(d,s) \rightarrow u(d,s) + \eta^{'}$.

As has already been discussed that 
spin-spin forces,  known to be compatible with the $\chi$QM
\cite{{riska},{chengspin},{prl}}, generate
configuration mixing \cite{{DGG},{Isgur},{yaouanc}}
which effectively leads to modification
of the spin polarization functions \cite{hd,su4}.
The most general configuration mixing generated by the spin-spin forces
in the case of octet baryons \cite{{Isgur},{yaouanc},{full}} 
can be expressed as
\be
|B \rangle=\left(|56,0^+\rangle_{N=0} \cos \theta +|56,0^+ \rangle_{N=2}  
\sin \theta \right) \cos \phi 
+  \left(|70,0^+\rangle_{N=2} \cos \theta^{'} +|70,2^+\rangle_{N=2}  
\sin \theta^{'} \right) \sin \phi\,, \label{full mixing}
\ee
where $\phi$ represents the $|56\rangle-|70\rangle$ mixing,
$\theta$ and  $\theta^{'}$ respectively correspond to the mixing 
among $|56,0^+\rangle_{N=0}-|56,0^+ \rangle_{N=2}$ states and
$|70,0^+\rangle_{N=2}-|70,2^+\rangle_{N=2}$ states.
For the present purpose, it is adequate
{\cite{{yaouanc},hd,{mgupta1}}} to consider the mixing only between
$|56,0^+ \rangle_{N=0}$ and the $|70,0^+\rangle_{N=2}$ states, for
example,
\begin{equation}
|B\rangle 
\equiv \left|8,{\frac{1}{2}}^+ \right> 
= \cos \phi |56,0^+\rangle_{N=0}
+ \sin \phi|70,0^+\rangle_{N=2}\,,  \label{mixed}
\end{equation} 
for details of the  spin, isospin and spatial parts of the 
wavefunction,  we  refer the reader to reference {\cite{{yaoubook}}. 
The mixing given above (Eq. (\ref{mixed})) would henceforth be referred to as 
configuration mixing.

To understand the partitioning of the nucleon spin in the 
$\chi$QM$_{gcm}$,  first we have to examine the various contributions
to the nucleon spin which arise in the context of $\chi$QM. In the
Cheng-Li version of $\chi$QM \cite{cheng}, at the leading order, the effective
degrees of freedom are constituent 
quarks and the weakly interacting GBs which fluctuate into the ``quark
 sea'', however the gluon contribution through the gluon anomaly 
has also been discussed \cite{{tommy},{chengspin}}. 
In the Adler-Bardeen scheme \cite{ab}, 
the flavor singlet component of
the spin of the proton is expressed as 
\be
\Delta \Sigma(Q^2)= \Delta \Sigma- 
 \frac{3\alpha_s(Q^2)}{2 \pi} \Delta g(Q^2)\,, \label{gluona}
\ee
where  $Q^2$ dependence on the right hand side is limited only to 
gluon polarization $\Delta
g(Q^2)$ and the strong coupling constant $\alpha_s(Q^2)$. 
As $\Delta \Sigma$ on the right hand side does not have any $Q^2$
dependence, therefore it can easily be calculated in the $\chi$QM. 
By considering $\Delta \Sigma(Q^2)$ experimental value at a
given $Q^2$ value as well as Eq. (\ref{gluona}), we can
calculate the contribution of the gluon polarization to the flavor
singlet component $\Delta g'(=\frac{3\alpha_s(Q^2)}{2 \pi} \Delta g(Q^2)$).
The above equation can also be expressed in terms of the quark spin
distribution functions as
\be
\Delta q(Q^2)= \Delta q- 
 \frac{\alpha_s(Q^2)}{2 \pi} \Delta g(Q^2)\,, \label{quarka}
\ee
where $\Delta q(Q^2)$ and $\Delta q$ are the experimentally measured and 
calculated quantities respectively.

Making use of Eq. (\ref{gluona}), Eq. (\ref{total spin}) can be
expressed as
\be
\frac {1}{2} =\frac{1}{2}\Delta \Sigma+\Delta L_{q}+\Delta L_{g}\,,  
\label{chiqm}
\ee
where
\be
\Delta \Sigma =\Delta \Sigma_{\rm val}+\Delta \Sigma_{\rm
sea}-\Delta g'\,, \label{sigma}
\ee
$\Delta L_{q}$ and $\Delta L_{g}$  
represent the angular momentum of the ``quark sea'' and gluons
respectively. 
Eqs. (\ref{chiqm}) and (\ref{sigma}), therefore provide us an
opportunity to calculate phenomenologically $\Delta g'$ and $\Delta
L_g$ within $\chi$QM by considering spin polarization functions and
spin dependent quantities.

In this
context, we first summarize the phenomenological quantities which can
be evaluated for fine tuning the $\chi$QM parameters as well as to
estimate the effect of the contribution of gluon polarization on the
spin polarization 
functions. To begin with, we
consider  $\Delta_3$ and $\Delta_8$ to be the appropriate spin
distribution functions for fitting the $\chi$QM parameters as these
are not only well known experimentally but also as discussed earlier,
have weak $Q^2$ dependence \cite{ellisq2}. 

The effect of the gluonic corrections on the spin distribution functions can
be estimated from Eq. (\ref{quarka}), however to incorporate the gluonic 
corrections on the magnetic moments,  we consider that the gluons remain 
part of the $q \bar q$ sea surrounding a given constituent quark as 
advocated by Cheng-Li \cite{cheng}. This effectively keeps the valence
and orbital contributions to the magnetic moments unchanged whereas the
contribution of the ``quark sea'' to the magnetic moment gets affected by the
presence of gluons.
The magnetic moment, including the effects of gluons, of a given baryon
which receives contributions from valence quarks, `` quark sea'' and its
orbital angular momentum is expressed as 
\cite{hdorbit}
\be
\mu(B)_{{\rm total}} = \mu(B)_{{\rm val}} + \mu(B)_{{\rm sea}} +
\mu(B)_{{\rm orbit}}\,.     \label{totalmag}
\ee
The valence contribution to the magnetic moment, in terms of quark spin
polarizations, can be written as 
\be
\mu(B)_{{\rm val}}=\sum_{q=u,d,s} {\Delta q_{{\rm val}}\mu_q}\,, 
 \label{mag}
\ee
where $\mu_q= \frac{e_q}{2 M_q}$ ($q=u,d,s$) is the quark magnetic moment,
$e_q$ and $M_q$ are the electric charge and the mass respectively for the
quark $q$. The valence quark spin polarization are as follows
\be
\Delta u_{{\rm val}} ={\cos}^2 \phi \left[\frac{4}{3} \right] 
   + {\sin}^2 \phi \left[\frac{2}{3}  \right],~~~
\Delta d_{{\rm val}} ={\cos}^2 \phi \left[-\frac{1}{3} \right]  +
  {\sin}^2 \phi \left[\frac{1}{3}  \right],~~~
\Delta s_{{\rm val}} = 0\,, \label{sigma1}
\ee 
giving 
\be
\Delta \Sigma_{\rm val}=\Delta u_{\rm val}+\Delta d_{\rm val}+
\Delta s_{\rm val}=1\,, \label{val}
\ee
and 
\be
 \mu(p)_{{\rm val}}=\left[{\cos}^2 \phi \left(\frac{4}{3} \right)
+{\sin}^2 \phi \left(\frac{2}{3} \right) \right ]\mu_u +
\left [{\cos}^2 \phi \left(-\frac{1}{3} \right)  +
  {\sin}^2 \phi \left(\frac{1}{3}  \right) \right]\mu_d +
[0]\mu_s\,. \label{val p} 
\ee
The sea contribution to the magnetic moment, in terms of the sea quark
spin polarizations, is 
\be 
\mu(B)_{{\rm sea}}=\sum_{q=u,d,s} {\Delta q_{{\rm sea}}\mu_q}\,.  
\label{magsea}
\ee
The sea quark spin polarizations in the presence of gluons can be
expressed as 
\be
{\Delta u}_{\rm sea}=-{\cos}^2 \phi \left[\frac{a}{3} (7+4 \alpha^2+
 \frac{4}{3}\beta^2 +\frac{8}{3} \zeta^2)\right]
-{\sin}^2 \phi \left[\frac{a}{3} (5+2 \alpha^2
+\frac{2}{3}\beta^2 +\frac{4}{3} \zeta^2)\right]- 
%\frac{\alpha_s(Q^2)}{2 \pi} 
\frac{1}{3}\Delta g'\,, \label{eqdu} 
\ee
\be
{\Delta d}_{\rm sea}=-{\cos}^2 \phi \left[\frac{a}{3} (2-\alpha^2
-\frac{1}{3}\beta^2 -\frac{2}{3} \zeta^2)\right]
-{\sin}^2 \phi \left[\frac{a}{3} (4+\alpha^2
+\frac{1}{3}\beta^2 +\frac{2}{3} \zeta^2)\right]- 
%\frac{\alpha_s(Q^2)}{2 \pi} 
\frac{1}{3}\Delta g'\,, \label{eqdd}
\ee
\be
 {\Delta s}_{\rm sea}=-a \alpha^2- 
%\frac{\alpha_s(Q^2)}{2 \pi} 
\frac{1}{3}\Delta g'\,. \label{eqds}
\ee
In terms of these the sea contribution to the magnetic moments can be
expressed through  Eq. (\ref{magsea}).  As already emphasized, we have to
consider the gluon modified 
$\Delta \Sigma'_{\rm sea}(=\Delta \Sigma_{\rm sea}-\Delta g'$) 
which in terms of 
 Eqs. (\ref{eqdu}), (\ref{eqdd}) and (\ref{eqds}) is given as
\be
{\Delta \Sigma'}_{\rm sea}={\Delta u}_{\rm sea}+{\Delta d}_{\rm
sea}+{\Delta s}_{\rm sea}=
 -\frac{a}{3}(9+6 \alpha^2+\beta^2+2 \zeta^2)-
%\frac{3 \alpha_s(Q^2)}{2 \pi} 
\Delta g'\,. 
\ee
We would like to add that
the above equation is also to be used
for studying the implications of gluonic contribution on the angular
momentum sum rule of the nucleon.

Following Ref. \cite{cheng1}, the total orbital 
angular momentum of the quark $q$ is given as
\be
\Delta L_{q}= \langle l_z \rangle \sum P_q \Delta \Sigma_{val}\,,
\ee
where $\sum P_q$ is the total transition probability of the quark $q$ 
and $\Delta \Sigma_{val}$ is the total spin carried by the valence quarks 
which is equal to one as calculated from Eq. (\ref{val}).
In the present context, the total angular momentum can be expressed in terms
of the $\chi$QM parameters as 

\be
\Delta L_{q}= \frac{a}{6}(9+6 \alpha^2+\beta^2+2 \zeta^2)\,. 
\label{ang}
\ee
Similarly, in the context of magnetic moments, the orbital angular momentum 
contribution of the sea, $\mu(B)_{{\rm orbit}}$, can be expressed in
terms of the valence   
quark polarizations and the orbital moments of the sea quarks, 
\be
\mu(p)_{{\rm orbit}} ={\cos}^2 \phi\left [ \frac{4}{3} 
[\mu (u_+ \rightarrow)] - \frac{1}{3} [\mu (d_+ \rightarrow)] \right]+
{\sin}^2 \phi\left[ \frac{2}{3} [\mu (u_+ \rightarrow)]
+\frac{1}{3} [\mu (d_+ \rightarrow)] \right ].  \label{orbit p} 
\ee
For details of the calculations of magnetic moments, we refer the
reader to Ref. \cite{hdorbit}.

For the sake of completeness, we have also calculated certain spin
dependent quantities which do not depend on gluon polarization,
nevertheless have implications for the parameters of $\chi$QM.
Some of the quantities are the weak axial-vector form factors
expressed as  
\bea
(G_A/G_V)_{n \rightarrow p}=\Delta_3 &=& \Delta u-\Delta d\,,\\
(G_A/G_V)_{\Lambda \rightarrow p} &=& \frac{1}{3}
(2 \Delta u-\Delta d-\Delta s)\,, \\
(G_A/G_V)_{\Sigma^- \rightarrow n} &=& \Delta d-\Delta s\,, \\
(G_A/G_V)_{\Xi^- \rightarrow \Lambda} &=& \frac{1}{3}
(\Delta u+\Delta d-2 \Delta s)\,. 
\eea

The  unpolarized valence quark distribution functions are not affected 
by configuration mixing, however a calculation of these quantities
also assumes importance in the present case as we attempt to effect a
unified fit to spin and quark distribution functions. 
As the $\chi$QM does not incorporate the $Q^2$ dependence,
therefore we consider only those quantities for our fit which are
independent of $Q^2$ or have weak $Q^2$ dependence.
The quark distribution functions which have
implications for the $\chi$QM parameters are the
antiquark flavor contents of the ``quark sea'' which
can be expressed as  \cite{{cheng},{song},{johan}} 
\be
\bar u =\frac{1}{12}[(2 \zeta+\beta+1)^2 +20] a\,,~~~
\bar d =\frac{1}{12}[(2 \zeta+ \beta -1)^2 +32] a\,, ~~~
\bar s =\frac{1}{3}[(\zeta -\beta)^2 +9 {\alpha}^{2}] a\,,
\ee
and
\be
u-\bar u=2\,, ~~~d-\bar d=1\,, ~~~ s-\bar s=0\,.
\ee
The deviation of
Gottfried  sum rule \cite{GSR}, expressed through Eq. (\ref{gsr}), can
be expressed in terms of the symmetry breaking parameters $\beta$ and
$\zeta$ as
\be
\left[I_G-\frac{1}{3}\right]=
\frac{2}{3}\left[\frac{a}{3}( 2 \zeta+ \beta-3)\right]. \label{zeta}
\ee
Similarly, $\bar u/\bar d$ {\cite{{baldit},{E866}}} measured 
through  the ratio of  muon pair production
cross sections  $\sigma_{pp}$ and $\sigma_{pn}$, is expressed in the 
present case as follows 
\be
\bar u/\bar d=\frac{(2 \zeta +\beta +1)^2+20}{(2 \zeta+ \beta-1)^2 +32}\,.
\ee
Some of the important quantities depending on the quark distribution functions which
are usually discussed in the literature are as follows
\bea
f_q&=&\frac{q+\bar q}{[\sum_{q} (q+\bar q)]}\,, \\
\frac{2 \bar s}{\bar u+ \bar d}&=& \frac{4[(\zeta-\beta)^2 +9\alpha^2]}
{(2\zeta+\beta)^2+27}\,, \\
\frac{2 \bar s}{u+d}&=& \frac{4a[(\zeta-\beta)^2 +9\alpha^2]}
{18+ a[(2\zeta+\beta)^2+27]}\,.
\eea

Before carrying out the detailed analysis involving quantities which
are dependent on $\Delta g'$, to begin with we have fixed the $\chi$QM
parameters using well determined quantities having weak $Q^2$ dependence, 
for example, $\Delta_3$, $\Delta_8$, $\bar u-\bar d$ asymmetry.
The $\chi$QM$_{gcm}$ invloves five parameters: $a$, $\alpha$,
$\beta$, $\zeta$ and $\phi$, the mixing angle $\phi$ is fixed from the
consideration of neutron charge radius as discussed earlier
\cite{{yaouanc},{full},{neu charge}}, whereas the pion fluctuation parameter
$a$ is also taken to be 0.1, in accordance with most of the other
calculations \cite{{song},{johan},{cheng1}}.
It has been shown \cite{{cheng},{johan},{hdorbit}} that to fix the
violation of Gottfried sum rule \cite{GSR}, we have to  consider the
relation
\be 
\bar u-\bar d=\frac{a}{3}(2 \zeta+\beta-3)\,.
\ee
In this relation, one immediately finds that for $a=0.1$, to reproduce 
$\bar u-\bar d$ asymmetry, one gets the relation $\zeta=-0.3-\beta/2$. 
The parameters $\alpha$ and $\beta$ are fixed by fitting $\Delta_3$
and $\Delta_8$. After having obtained these parameters, we have also
considered the variation in these parameters to obtain the best fit
corresponding to quark distribution functions and spin polarization
functions. Interestingly, the value of these parameters found above
give the best fit. In Table \ref{input}, we summarize the input 
parameters and their values.
In Table \ref{spininput}, we have presented the phenomenological
quantities which do
not depend on $\Delta g'$ and have been used in fitting the $\chi$QM
parameters. In the table, we have presented the results both with and
without configuration mixing and it can be seen that configuration
mixing is very much needed to obtain the fit and it improves the
results in right direction, however the variation in the mixing angle
does not lead to much improvement.

After having fixed these $\chi$QM parameters, we have calculated the
phenomenological quantities depending on the quark distribution
functions having implications for these and the 
corresponding results have been presented in 
Table \ref {quark}.
These parameters do not have dependence on $\Delta g'$ as well as the
mixing angle, however these have been presented for the sake of
completeness as well as to give an idea of the unified fit obtained
within $\chi$QM$_{gcm}$ for the quark distribution functions and spin
polarization functions. Without getting into the
detailed discussion, we would just like to mention that a cursory look
at the table shows that we are able to obtain an 
excellent fit to the flavor distribution functions, for example,
$\bar u-\bar d$,  $\bar u/\bar d$, $\frac{2 \bar s}{\bar u+\bar d}$,  
$\frac{2 \bar s}{u+d}$, $f_s$ and $f_3/f_8$. 

In Table \ref{spin}, we have presented the various phenomenological
quantities which are dependent on $\Delta g'$. 
To study the contribution of gluon polarization in the
partitioning of the nucleon spin and its implications on the other
quantities, we need to calculate the value of $\Delta g'$. 
Using $\Delta \Sigma=0.62$, obtained from Ref. \cite{hd}, 
as well as the experimental values 
$\Delta \Sigma(Q^2=5{\rm GeV}^2)=0.30\pm0.06$ \cite{abe} and  
$\alpha_s(Q^2=5{\rm GeV}^2)= 0.287\pm0.020$ \cite{PDG}, 
from Eq. (\ref{gluona}) $\Delta g'$ comes out to be 0.33.
In the table we have also presented the results of the calculations carried
out with different values of mixing angle, again to highlight the fact
that the value of mixing angle considered earlier almost gives the
best fit.
A general look at the table shows that the results of all 
the quantities affected by the inclusion of gluon polarization get 
improved in the right direction.
It is interesting to note that the inclusion of $\Delta g'$ improves 
the results in the right direction even when configuration mixing is
not included, however, when configuration mixing is included, these
show considerable further improvement. 
In fact, the results are almost perfect
for the spin polarization functions after the inclusion of $\Delta g'$,
the magnitude of $\Delta u$ decreases whereas the
magnitude of $\Delta d$ increases giving an almost perfect fit. 

The improvement in the spin polarization functions after the inclusion 
of the effect of gluon polarization suggests corresponding improvements in the
magnetic moments also, however, for the better appreciation of the
role of gluon polarization in magnetic moments, it is desirable to
discuss very briefly the key ingredients of Ref. \cite{hdorbit} wherein
an excellent agreement had been achieved for the baryon magnetic
moments as well as for the violation in the Coleman-Glashow sum rule
\cite{cg}. It had been discussed in detail there that the Cheng-Li
mechanism \cite{cheng1}, incorporating the sea quark polarization and
the orbital 
angular momentum, as well as configuration mixing are the key
ingredients for registering the agreement achieved in Ref.
\cite{hdorbit}. A detailed scrutiny of the results however reveals that
there are still discrepancies  of the order of 5\% compared with the
experimental results. 
Interestingly, when the contribution of gluon polarization, found from
Eq. (\ref{gluona}) by fitting $\Delta \Sigma$, is incorporated in the
magnetic moments as shown in  Eqs. (\ref{eqdu}), (\ref{eqdd}) and 
(\ref{eqds}), we find further improvements in Ref. \cite{hdorbit}.
For example, in the case of $\mu_p$, 
$\mu_{\Sigma^-}$, $\mu_{\Xi^0}$, $\mu_{\Lambda}$ and 
$\mu_{\Sigma \Lambda}$ hardly anything is left to desire, whereas in
the case of $\mu_n$ and $\mu_{\Sigma^+}$, the discrepancy is less than
5\%. In the case of $\mu_n$, $\mu_{\Sigma^-}$ and $\mu_{\Sigma
\Lambda}$, the results get reduced in the right direction by the 
inclusion of $\Delta g'$ giving a much better fit. On the 
other hand, $\mu_{\Xi^-}$ and $\mu_{\Lambda}$ increase after the inclusion 
of gluon polarization again giving a better overlap with the data. 
It needs to be mentioned that even in the case of $\mu_{\Xi^-}$,
a difficult case for most of the models, the inclusion of corrections due 
to gluon polarization leads to a considerable improvement. This strongly 
suggests that our phenomenological evaluation of $\Delta g'$ seems to be 
in the right direction with constituent quarks, weakly interacting GBs and
weakly interacting gluons constituting the appropriate degrees
of freedom in the nonperturbative regime as already emphasized in 
Ref. \cite{hdorbit} which is also in agreement with the basic tenets of 
chiral quark soliton model \cite{cqsm}.
One can also examine the effects of gluon polarization on the decuplet 
magnetic moments, however the effects are only 
marginal in this case.
%It may be of interet to mention here that $\Delta g$, as it occurs in
%$\Delta g'$, comes out to be 2.33 which is in fair 
%agreement with certain recent measurements \cite{hermes} 
%as well as theoretical estimates \cite{{deltag},{bag}}. 

After having examined the implications of $\Delta g'$ for magnetic
moments, one would like to study its role in understanding the
angular momentum sum rule of the proton within $\chi$QM. 
A scrutiny of  Table \ref{total}, where we have considered the values 
of various contributions to the proton angular momentum sum rule, 
reveals several interesting points.  In case
the sum rule is to be explained in terms of the spin
polarizations of valence quarks, polarizations as well as the angular
momentum of the ``quark sea'' and the gluon polarization, then these
should add on to give the total spin of the nucleon. 
In case there is a discrepancy then the balance has to be attributed
to the gluon angular momentum  which cannot be calculated directly 
in the present context.
Without the gluon angular momentum, we find that the nucleon angular
momentum falls short by 0.16 from the $\frac{1}{2}$ value, therefore
we tend to assign this value to the gluon angular momentum which is 
almost equal to the angular momentum of the ``quark sea''. At present,
we do not have any deep understanding of these values, however they do
indicate that these contributions may not be negligible even in a more
rigorous model.

To summarize, apart from carrying out the detailed analysis of the
$\chi$QM with configuration mixing with the latest data pertaining to 
$\bar u-\bar d$ asymmetry and the spin polarization functions, we have
phenomenologically evaluated the contribution of gluon polarization to
the flavor singlet component of the total spin through the relation 
$\Delta\Sigma(Q^2)=\Delta\Sigma-\frac{3\alpha_s(Q^2)}{2\pi}\Delta g(Q^2)$
as defined by the Adler-Bardeen scheme.
When the effects of gluon polarization are included in the case of quark spin 
polarization functions and baryon magnetic moments, we obtain an
almost perfect fit for the quark spin polarization functions and many of the 
baryon magnetic moments with good deal of improvement in the other cases. 
In case an attempt is made to explain the angular momentum sum rule for 
proton with the inclusion of gluon polarization effects, we find 
the contribution of gluon angular momentum to be as important as that of
the $q \bar q$ pairs.

\vskip .2cm
 {\bf ACKNOWLEDGMENTS}\\
The authors would like to thank S.D. Sharma, M. Randhawa and J.M.S. Rana
for a few useful discussions.
H.D. would like to thank CSIR, Govt. of India, for
 financial support and the chairman,
 Department of Physics, for providing facilities to work
 in the department.

\begin{table}
\begin{center}
\begin{tabular}{|cccccc|}      \hline
Parameter &$\phi$ &$a$ &$\alpha$ &$\beta$ &$\zeta$ \\  
\hline
Value & $20^{o}$ & 0.1 & 0.4 & 0.7 & $-0.3-\beta/2$  \\ \hline

\end{tabular}
\end{center}
\caption{ Input parameters and their values used in the analysis.} 
\label{input}
\end{table}

\begin{table}
\begin{center}
\begin{tabular}{|ccccccc|}       \hline
Parameter & Data & NRQM & $\chi$QM  &  \multicolumn{3}{c|}{$\chi$QM$_{gcm}$} 
\\ \cline{5-7} 
& &  &   & $\phi=16^{o}$ & $\phi=18^{o}$ & $\phi=20^{o}$ \\  \hline

$\Delta_8$ & 0.58  $\pm$ .025 {\cite{PDG}}& 1 &0.67 & 0.67& 0.67& 0.67 \\ 

$\Delta_3=(G_A/G_V)_{n \rightarrow p}$ & 1.26 $\pm$ 0.0035 \cite{PDG}& 1.66 &
1.40 & 1.30 & 1.28& 1.26 \\
$(G_A/G_V)_{\Lambda \rightarrow p}$ & 0.72 $\pm$ 0.02 \cite{PDG} & 1 & 
 0.81 &0.76 & 0.75 & 0.74   \\
$(G_A/G_V)_{\Sigma^- \rightarrow n}$ & $-$0.34 $\pm$ 0.02 \cite{PDG}& $-$0.33 &
$-$0.36 & $-$0.32 & $-$0.31 & $-$0.30  \\
$(G_A/G_V)_{\Xi^- \rightarrow \Lambda}$ & 0.25 $\pm$ 0.05 \cite{PDG} & 0.33&
 0.22 &0.22 & 0.22 & 0.22  \\ \hline
\end{tabular}
\end{center}
\caption{The phenomenological values of the spin dependent quantities
used for fitting the $\chi$QM input parameters. } \label{spininput}
\end{table}

\begin{table}
\begin{center}
\begin{tabular}{|cccc|}       \hline
              
Parameter & Data & NRQM & $\chi$QM \\ \hline
$\bar u$ &-  &- &  0.21  \\

$\bar d$ & - &- &  0.33  \\

$\bar s$ &-  &- &   0.10  \\

$\bar u-\bar d$ & $-0.118 \pm$ .015 \cite{E866} &0 & $-0.12$  \\

$\bar u/\bar d$ & 0.67 $\pm$ 0.06 {\cite{E866}} & 1 & 0.63   \\

$I_G$ & 0.266  $\pm$ .005 & 0.33 & 0.253   \\
    
$\frac{2 \bar s}{\bar u+ \bar d}$ & 0.477 $\pm$ .051 {\cite{ao}} 
&- &  0.41 \\

$\frac{2 \bar s}{u+d}$ & 0.099 $\pm$ .009 {\cite{ao}} & - & 0.06   \\

$f_u$ &-  &- &   0.56  \\

$f_d$ &- & -&  0.39  \\

$f_s$ &  0.10 $\pm$ 0.06 {\cite{ao}} & 0 &  0.06   \\

$f_3=$  & -&- & 0.18 \\
$f_u-f_d$  &  & &   \\

$f_8=$ &- &- & 0.86 \\
$f_u+f_d-2 f_s$ &   & &  \\

$f_3/f_8$ & 0.21 $\pm$ 0.05 {\cite{cheng}} & 0.33 &  0.21  \\ \hline

\end{tabular}
\end{center}
\caption{The flavor dependent spin independent quark distribution 
functions
not affected by the inclusion of configuration mixing and the
gluon polarization.}  
\label{quark}
\end{table}

\begin{table}
\tabcolsep 0.2mm
%{\footnotesize
\begin{center}
\begin{tabular}{|ccc|cccc|cccc|}       \hline
 & & & \multicolumn{4}{|c|} {Without $\Delta g'$} &
\multicolumn{4}{c|} {With $\Delta g'$}\\ \cline{4-11} 
Parameter & Data & NRQM & $\chi$QM  &  \multicolumn{3}{c|} {$\chi$QM$_{gcm}$} & 
$\chi$QM  &  \multicolumn{3}{c|} {$\chi$QM$_{gcm}$} \\ 
&&&& $\phi=16^{o}$ & $\phi=18^{o}$ & $\phi=20^{o}$ & &$\phi=16^{o}$ & $\phi=18^{o}$& $\phi=20^{o}$ \\  \hline
$\Delta u$ & 0.85 $\pm$ 0.05 \cite{adams} & 1.33 & 1.02 &0.97 & 0.96 & 0.95  & 
0.91  & 0.82 & 0.83 & 0.84   \\ 

$\Delta d$ & $-$0.41  $\pm$ 0.05 \cite{adams}  & $-0.33$ & $-$0.38 &
$-$0.33 & $-$0.32 & $-$0.31  & $-$0.49 & $-$0.44 &$-$0.43 & $-$0.42  \\

$\Delta s$ &$-$0.07  $\pm$ 0.05 \cite{adams} & 0  &$-$0.02 &$-$0.02 & $-$0.02&  $-$0.02
  & $-$0.13& $-$0.13& $-$0.13& $-$0.13 \\  \hline

$\mu_p$ & 2.79$\pm$0.00 \cite{PDG}& 2.72  & 3.03  &2.81 & 2.80&  2.80  & 
3.00 &2.78 & 2.77 & 2.77 \\ 
$\mu_n$ & $-$1.91$\pm$0.00 \cite{PDG}& $-$1.81 & $-$2.24 & $-$2.00& $-$1.99& $-$1.99 & 
$-$2.21 & $-$1.97 & $-$1.96 & $-$1.96 \\ 
$\mu_{\Sigma^-}$ & $-$1.16$\pm$0.025\cite{PDG} &  $-$1.01  & $-$1.26 & $-$1.21& $-$1.20& $-$1.20 & $-$1.23 & $-$1.18 & $-$1.17 & $-$1.17 \\
$\mu_{\Sigma^+}$ & 2.45$\pm$0.01\cite{PDG}  &  2.61& 2.62  & 2.44 & 2.43 & 2.43& 2.59& 2.41 & 2.40 & 2.40 \\
$\mu_{\Xi^0}$    & $-$1.25$\pm$0.014\cite{PDG} & $-$1.41 & $-$1.47 & $-$1.25& $-$1.24& $-$1.24 & $-$1.50 & $-$1.28 & $-$1.27 & $-$1.27 \\
$\mu_{\Xi^-}$    & $-$0.65$\pm$0.002\cite{PDG} & $-$0.50 & $-$0.54  & $-$0.55 & $-$0.56 & $-$0.56 & $-$0.57& $-$0.58 & $-$0.59  & $-$0.59 \\
$\mu_{\Lambda}$ & $-$0.61$\pm$0.004\cite{PDG} & $-$0.59 & $-$0.68  & $-$0.58 & $-$0.59 & $-$0.59 & $-$0.71 & $-$0.63 & $-$0.62 & $-$0.62 \\ 
$\mu_{\Sigma \Lambda}$ & 1.61$\pm$0.08\cite{PDG} & 1.51 & 1.72 & 1.62 & 1.63 & 1.63 & 1.69 & 1.61 & 1.60 & 1.60 \\ \hline

\end{tabular}
\end{center}
\caption{The effect of contribution of gluon polarization on the quark 
spin polarization functions and the octet baryon magnetic moments
in the $\chi$QM$_{gcm}$. } \label{spin}
\end{table}

\begin{table}
\begin{center}
\begin{tabular}{|ccccc|}       \hline
Parameter  &Data & NRQM & $\chi$QM$_{gcm}$ without $\Delta g'$ 
& $\chi$QM$_{gcm}$ with $\Delta g'$    \\ \hline
$\Delta \Sigma$ &0.30$\pm0.05$ \cite{abe} & 1  & 0.62  & 0.30 \\
$\Delta L_{q}$ & $$-$$ &0 & 0.19 & 0.19  \\ 
$\Delta L_{g}$ & $$-$$ &0 & 0 & 0.16  \\ \hline
$\frac {1}{2}\Delta \Sigma+\Delta L_{q}+\Delta L_{g}$ & 0.5 & 0.5 & 0.5 & 0.5  \\
\hline
\end{tabular}
\end{center}
\caption{The contributions of various terms to the angular momentum
sum rule.} \label{total}
\end{table}

\end{document}